\documentclass[12pt]{article}
\usepackage{amsmath}
\usepackage{amssymb}

\begin{document}

\begin{center}
\textbf{2d GRAVITY WITH TORSION, ORIENTED MATROIDS}

\smallskip\ 

\textbf{AND 2+2 DIMENSIONS}

\smallskip\ 

J. A. Nieto$^{\star \ast }$ \footnote{%
nieto@uas.uasnet.mx} and E. A. Le\'{o}n$^{\ast }$ \footnote{%
ealeon@posgrado.cifus.uson.mx}

\smallskip\ 

$^{\star }$\textit{Facultad de Ciencias F\'{\i}sico-Matem\'{a}ticas de la
Universidad Aut\'{o}noma} \textit{de Sinaloa, 80010, Culiac\'{a}n Sinaloa, M%
\'{e}xico.}

$^{\ast }$\textit{Departameto de Investigaci\'{o}n en F\'{\i}sica de la
Universidad de Sonora, 83000, Hermosillo Sonora , M\'{e}xico}

\bigskip\ 

\bigskip\ 

\textbf{Abstract}
\end{center}

We find a link between oriented matroid theory and 2d gravity with torsion.
Our considerations may be useful in the context of noncommutative phase
space in a target spacetime of signature (2+2) and in a possible theory of
gravity ramification.

\bigskip\ 

\bigskip\ 

\bigskip\ 

\bigskip\ 

\bigskip\ 

\bigskip\ 

\bigskip\ 

\bigskip\ 


Pacs numbers: 04.60.-m, 04.65.+e, 11.15.-q, 11.30.Ly

September, 2009

\newpage

As it is known, the theory of matroids is a fascinating topic in mathematics
[1]. Why should not be also interesting in some scenarios of physics? We are
convinced that matroid theory should be an essential part not only of
physics in general, but also of M-theory. In fact, it seems that the duality
concept that brought matroid theory from a matrix formalism in 1935, with
the work of Whitney (see Ref. [2] and references therein), is closely
related to the duality concept that brought M-theory from string theory in
1994 (see Refs. [3-11] for connections between matroids and various subjects
of high energy physics and gravity). These observations are some of the main
motivations for the proposal [12] of considering oriented matroid theory as
the mathematical framework for M-theory. In this paper, we would like to
report new progress in our quest of connecting matroid theory with different
scenarios of high energy physics and gravity. Specifically, we find a
connection between matroids and $2d$ gravity with torsion and $2+2$
dimensions. In the route we find many new directions in which one can pursue
further research, such as tame and wild ramification [13], nonsymmetric
gravitational theory (see Ref. [14] and references therein) and Clifford
algebras (see Ref. [15] and references therein). We believe that our results
may be of particular interest not only for physicists but also for
mathematicians.

In order to achieve our goal we first show that a $2\times 2$-matrix
function in two dimensions can be interpreted in terms of a metric
associated with 2d gravity with torsion. Let us start by writing a complex
number $z$ in the traditional form [16]

\begin{equation}
z=x+iy,  \label{1}
\end{equation}%
where $x$ and $y$ are real numbers and $i^{2}=-1$. There exist, however,
another, less used, way to write a complex number, namely [17]

\begin{equation}
\left( 
\begin{array}{cc}
x & y \\ 
-y & x%
\end{array}%
\right) .  \label{2}
\end{equation}%
In this case the product of two complex numbers corresponds to the usual
matrix product. These two representations of complex numbers can be linked
by writing (2) as

\begin{equation}
\left( 
\begin{array}{cc}
x & y \\ 
-y & x%
\end{array}%
\right) =x\left( 
\begin{array}{cc}
1 & 0 \\ 
0 & 1%
\end{array}%
\right) +y\left( 
\begin{array}{cc}
0 & 1 \\ 
-1 & 0%
\end{array}%
\right) .  \label{3}
\end{equation}%
Since $\left( 
\begin{array}{cc}
0 & 1 \\ 
-1 & 0%
\end{array}%
\right) \left( 
\begin{array}{cc}
0 & 1 \\ 
-1 & 0%
\end{array}%
\right) =-\left( 
\begin{array}{cc}
1 & 0 \\ 
0 & 1%
\end{array}%
\right) $ one finds from (1) and (3) that the matrix $\left( 
\begin{array}{cc}
0 & 1 \\ 
-1 & 0%
\end{array}%
\right) $ can be identified with the imaginary unit $i$.

It turns out that the matrices $\left( 
\begin{array}{cc}
1 & 0 \\ 
0 & 1%
\end{array}%
\right) $ and $\left( 
\begin{array}{cc}
0 & 1 \\ 
-1 & 0%
\end{array}%
\right) $ can be considered as two of the matrix bases of general real $%
2\times 2$ matrices which we denote by $M(2,R)$. In fact, any $2\times 2$
matrix $\Omega $ over the real can be written as

\begin{equation}
\begin{array}{c}
\Omega =\left( 
\begin{array}{cc}
a & b \\ 
c & d%
\end{array}%
\right) =x\left( 
\begin{array}{cc}
1 & 0 \\ 
0 & 1%
\end{array}%
\right) +y\left( 
\begin{array}{cc}
0 & 1 \\ 
-1 & 0%
\end{array}%
\right) \\ 
\\ 
+r\left( 
\begin{array}{cc}
1 & 0 \\ 
0 & -1%
\end{array}%
\right) +s\left( 
\begin{array}{cc}
0 & 1 \\ 
1 & 0%
\end{array}%
\right) ,%
\end{array}
\label{4}
\end{equation}%
where

\begin{equation}
\begin{array}{ccc}
x=\frac{1}{2}(a+d), &  & y=\frac{1}{2}(b-c), \\ 
&  &  \\ 
r=\frac{1}{2}(a-d), &  & s=\frac{1}{2}(b+c).%
\end{array}
\label{5}
\end{equation}

Let us rewrite (4) in the form

\begin{equation}
\Omega _{ij}=x\delta _{ij}+y\varepsilon _{ij}+r\eta _{ij}+s\lambda _{ij},
\label{6}
\end{equation}%
where

\begin{equation}
\begin{array}{ccc}
\delta _{ij}\equiv \left( 
\begin{array}{cc}
1 & 0 \\ 
0 & 1%
\end{array}%
\right) , &  & \varepsilon _{ij}\equiv \left( 
\begin{array}{cc}
0 & 1 \\ 
-1 & 0%
\end{array}%
\right) , \\ 
&  &  \\ 
\eta _{ij}\equiv \left( 
\begin{array}{cc}
1 & 0 \\ 
0 & -1%
\end{array}%
\right) , &  & \lambda _{ij}\equiv \left( 
\begin{array}{cc}
0 & 1 \\ 
1 & 0%
\end{array}%
\right) .%
\end{array}
\label{7}
\end{equation}%
Considering this notation, we find that (1) becomes

\begin{equation}
z_{ij}=x\delta _{ij}+y\varepsilon _{ij}.  \label{8}
\end{equation}%
Comparing (6) and (8), we see that (8) can be obtained from (6) by setting $%
r=0$ and $s=0.$ If $ad-bc\neq 0,$ that is if $\det \Omega \neq 0$, then the
matrices in $M(2,R)$ can be associated with the group $GL(2,R)$. If we
further require $ad-bc=1$, then one gets the elements of the subgroup $%
SL(2,R)$. It turns out that this subgroup is of special interest in 2t
physics [18-20].

Now, consider the following four functions $F(x,y,r,s),G(x,y,r,s),H(x,y,r,s)$
and $Q(x,y,r,s)$, and construct the matrix

\begin{equation}
\gamma =\left( 
\begin{array}{cc}
F & G \\ 
H & Q%
\end{array}%
\right) .  \label{9}
\end{equation}%
By setting

\begin{equation}
\begin{array}{ccc}
u=\frac{1}{2}(F+Q), &  & v=\frac{1}{2}(G-H), \\ 
&  &  \\ 
w=\frac{1}{2}(F-Q), &  & \xi =\frac{1}{2}(G+H),%
\end{array}
\label{10}
\end{equation}%
we get that $\gamma $ can be written as

\begin{equation}
\gamma =u\left( 
\begin{array}{cc}
1 & 0 \\ 
0 & 1%
\end{array}%
\right) +v\left( 
\begin{array}{cc}
0 & 1 \\ 
-1 & 0%
\end{array}%
\right) +w\left( 
\begin{array}{cc}
1 & 0 \\ 
0 & -1%
\end{array}%
\right) +\xi \left( 
\begin{array}{cc}
0 & 1 \\ 
1 & 0%
\end{array}%
\right) ,  \label{11}
\end{equation}%
or

\begin{equation}
\gamma _{ij}=u\delta _{ij}+v\varepsilon _{ij}+w\eta _{ij}+\xi \lambda _{ij}.
\label{12}
\end{equation}%
We can always decompose the matrix $\gamma _{ij}$ in terms of its symmetric $%
g_{ij}=g_{ji}$ and antisymmetric $A_{ij}=-A_{ji}$ parts. In fact, we have

\begin{equation}
\gamma _{ij}(x,y,r,s)=g_{ij}(x,y,r,s)+A_{ij}(x,y,r,s).  \label{13}
\end{equation}%
From (11) or (12) we find that we can write $g_{ij}(x,y,r,s)$ in form

\begin{equation}
\begin{array}{c}
g_{ij}(x,y,r,s)=u(x,y,r,s)\left( 
\begin{array}{cc}
1 & 0 \\ 
0 & 1%
\end{array}%
\right) +w(x,y,r,s)\left( 
\begin{array}{cc}
1 & 0 \\ 
0 & -1%
\end{array}%
\right) \\ 
\\ 
+\xi (x,y,r,s)\left( 
\begin{array}{cc}
0 & 1 \\ 
1 & 0%
\end{array}%
\right) ,%
\end{array}
\label{14}
\end{equation}%
while

\begin{equation}
A_{ij}(x,y,r,s)=v(x,y,r,s)\left( 
\begin{array}{cc}
0 & 1 \\ 
-1 & 0%
\end{array}%
\right) .  \label{15}
\end{equation}

An interesting possibility emerges by dimensional reduction of the variables 
$r$ and $s$, that is by setting in (13) $r=0$ and $s=0$. We have

\begin{equation}
\gamma _{ij}(x,y)=g_{ij}(x,y)+A_{ij}(x,y),  \label{16}
\end{equation}%
with

\begin{equation}
g_{ij}(x,y)=u(x,y)\delta _{ij}+w(x,y)\eta _{ij}+\xi (x,y)\lambda _{ij}
\label{17}
\end{equation}%
and%
\begin{equation}
A_{ij}(x,y)=v(x,y)\varepsilon _{ij}.  \label{18}
\end{equation}%
Of course, according to (8) the expressions (16), (17) and (18) can be
associated with a complex structure. This observation can be clarified by
using isothermal coordinates in which $w=0$ and $\xi =0$. In this case (16)
is reduced to

\begin{equation}
f_{ij}(x,y)=u(x,y)\delta _{ij}+v(x,y)\varepsilon _{ij},  \label{19}
\end{equation}%
where we wrote $\gamma _{ij}(x,y)\rightarrow f_{ij}(x,y)$ in order to
emphasize this reduction. In the traditional notation, (19) becomes $%
f(x,y)=u(x,y)+iv(x,y)$. It turns out that the existence of isothermal
coordinates is linked to the Cauchy-Riemann conditions for $u$ and $v$,
namely $\partial u/\partial x=\partial v/\partial y$ and $\partial
u/\partial y=-\partial v/\partial x$ [16].

One of the main reason for the above discussion comes from the question: is
it possible to identify the symmetric matrix $g_{ij}(x,y)$ with 2d gravity?
Assuming that this is the case the next question is then: what kind of
gravitational theory describes $\gamma _{ij}(x,y)$? In what follows we shall
show that $\gamma _{ij}(x,y)$ can be identified not only with a nonsymmetric
gravitational theory in two dimensions but also with 2d gravity with
torsion. First, consider the covariant derivative of the metric tensor

\begin{equation}
\bigtriangledown _{k}g_{ij}=\partial _{k}g_{ij}-\Gamma
_{ki}^{l}g_{lj}-\Gamma _{kj}^{l}g_{il}=0.  \label{20}
\end{equation}%
Here, we assume that the symbols $\Gamma _{ki}^{l}$ are not necessarily
symmetric in the two indices $k$ and $i$. In fact, if we define the torsion
as $T_{ki}^{l}\equiv \Gamma _{ki}^{l}-\Gamma _{ik}^{l}$, one finds that the
more general solution of (20) is

\begin{equation}
\Gamma _{kij}=\frac{1}{2}(\partial _{k}g_{ji}+\partial _{i}g_{jk}-\partial
_{j}g_{ki})-\frac{1}{2}(T_{kji}+T_{ijk}-T_{kij}),  \label{21}
\end{equation}%
where $\Gamma _{kij}=\Gamma _{ki}^{l}g_{lj}$ and $T_{kij}=T_{ki}^{l}g_{lj}$.

On the other hand, if we consider the expression

\begin{equation}
\frac{1}{2}(\partial _{k}\gamma _{ji}+\partial _{i}\gamma _{jk}-\partial
_{j}\gamma _{ik}),  \label{22}
\end{equation}%
by substituting (16) into (22) one gets

\begin{equation}
\frac{1}{2}(\partial _{k}\gamma _{ji}+\partial _{i}\gamma _{jk}-\partial
_{j}\gamma _{ik})=\frac{1}{2}(\partial _{k}g_{ji}+\partial
_{i}g_{jk}-\partial _{j}g_{ki})+\frac{1}{2}(\partial _{k}A_{ji}+\partial
_{i}A_{jk}-\partial _{j}A_{ik}).  \label{23}
\end{equation}%
Comparing (23) and (21) one sees that if one sets $T_{kji}=\partial
_{i}A_{kj}$ the expression (23) can be identified with the connection $%
\Gamma _{kij}$ which presumably describes 2d gravity with torsion. Since $%
A_{ij}$ can always be written as (18) we discover that (23) yields

\begin{equation}
\Gamma _{kij}=\frac{1}{2}(\partial _{k}g_{ji}+\partial _{i}g_{jk}-\partial
_{j}g_{ki})+\frac{1}{2}(v_{,k}\varepsilon _{ji}+v_{,i}\varepsilon
_{jk}-v_{,j}\varepsilon _{ik}).  \label{24}
\end{equation}%
Here, we used the notation $\partial _{k}v=v_{,k}$.

The curvature Riemann tensor can be defined as usual

\begin{equation}
\mathcal{R}_{kij}^{m}=\partial _{i}\Gamma _{kj}^{m}-\partial _{j}\Gamma
_{ki}^{m}+\Gamma _{ni}^{m}\Gamma _{kj}^{n}-\Gamma _{nj}^{m}\Gamma _{ki}^{n}.
\label{25}
\end{equation}%
The proposed gravitational theory, which may be interesting in string
theory, can have a density Lagrangian $\mathcal{L}$ of the form $\mathcal{%
L\sim R}^{2}+\Lambda $ [21], where $\Lambda $ is a constant. In this
context, we have proved that it makes sense to consider the nonsymmetric
metric of the form (16)-(18) as a 2d gravity with torsion.

From the point of view of complex structure there are a number of
interesting issues that arises from the above formalism. One may be
interested, for instance, in considering the true degrees of freedom for the
metric $g_{ij}$. In this case one may start with the Teichmuller space
associated with the metric $g_{ij}$ and then to determine the Moduli space
of such a metric [22]. Another possibility is to consider similarities. In
this case one may be interested to associate with the metric $g_{ji}$ either
the Riemann-Roch theorem [23] or the tame and wild ramification complex
structure [13]. In the later case one may assume that the principal part of
the metric $g_{ij}$ looks like%
\begin{equation}
g_{ij}(x,y)=(\frac{T_{n}}{z^{n}}+\frac{T_{n-1}}{z^{n-1}}+...+\frac{T_{1}}{z}%
)\delta _{ij}.  \label{26}
\end{equation}%
In this case the similarities can be identified with solitons associated
with black holes. In this scenario our constructed route to 2d gravity with
torsion provides a bridge which may bring many ideas from complex structure
to 2d gravity with torsion and \textit{vice versa}.

Let us now study some aspects of the above formalism from the point of view
of matroid theory. Consider the matrix

\begin{equation}
V_{i}^{A}=\left( 
\begin{array}{cccc}
1 & 0 & 0 & 1 \\ 
0 & 1 & -1 & 0%
\end{array}%
\right) ,  \label{27}
\end{equation}%
with the index $A$ taking values in the set 
\begin{equation}
E=\{1,2,3,4\}.  \label{28}
\end{equation}%
It turns out that the subsets $\{\mathbf{V}^{1},\mathbf{V}^{2}\}$, $\{%
\mathbf{V}^{1},\mathbf{V}^{3}\}$, $\{\mathbf{V}^{2},\mathbf{V}^{4}\}$ and $\{%
\mathbf{V}^{3},\mathbf{V}^{4}\}$ are bases over the real of the matrix (27).
One can associate with these subsets the collection%
\begin{equation}
\mathcal{B}=\{\{1,2\},\{1,3\},\{2,4\},\{3,4\}\},  \label{29}
\end{equation}%
which can be understood as a family of subsets of $E$. It is not difficult
to show that the pair $\mathcal{M}=(E,\mathcal{B})$ is a 2-rank self-dual
matroid. The fact that we can express $\mathcal{M}$ in the matrix form (27)
means that this matroid is representable (or realizable) [1]. Moreover, one
can show that this matroid is graphic and orientable. In the later case the
corresponding chirotope [1] is given by

\begin{equation}
\chi ^{AB}=\varepsilon ^{ij}V_{i}^{A}V_{j}^{B}.  \label{30}
\end{equation}%
Thus, we get, as nonvanishing elements of the chirotope $\chi ^{AB}$, the
combinations

\begin{equation}
\begin{array}{cccc}
12+, & 13-, & 24-, & 34+.%
\end{array}
\label{31}
\end{equation}%
The relation of this matroid structure with of our previous discussion comes
from the identification $\{\mathbf{V}^{1},\mathbf{V}^{2}\}\rightarrow \delta
_{ij}$, $\{\mathbf{V}^{1},\mathbf{V}^{3}\}\rightarrow \eta _{ij}$, $\{%
\mathbf{V}^{2},\mathbf{V}^{4}\}\rightarrow \lambda _{ij}$ and $\{\mathbf{V}%
^{3},\mathbf{V}^{4}\}\rightarrow \varepsilon _{ij}$. The signs in (31)
correspond to the determinants of the matrices $\delta _{ij}$, $\eta _{ij}$, 
$\lambda _{ij}$ and $\varepsilon _{ij}$, which can be calculated using (30).
Therefore, what we have shown is that the bases of $M(2,R)$ as given in (4)
(or (7)) admit an oriented matroid interpretation. It may be of some
interest to consider the weak mapping $\mathcal{B\rightarrow B}_{c}$ with%
\begin{equation}
\mathcal{B}_{c}=\{\{1,2\},\{3,4\}\},  \label{32}
\end{equation}%
leading to the reduced pair $\mathcal{M}_{c}=(E,\mathcal{B}_{c})$. When the
local structure is considered as in (14)-(18), one needs to rely in the
matroid fiber bundle notion (see Refs. [24] and [25] and references
therein). Therefore, we have found a link which connect matroid fiber bundle
with 2d gravity with torsion.

It is worth mentioning the following observations. It is known that the
fundamental matrices $\delta _{ij},\eta _{ij},\lambda _{ij}$ and $%
\varepsilon _{ij}$ given in (7) not only form a basis for $M(2,R)$ but also
determine a basis for the Clifford algebras $C(2,0)$ and $C(1,1)$. In fact
one has the isomorphisms $M(2,R)\sim C(2,0)\sim C(1,1)$. Moreover, one can
show that $C(0,2)$ can be constructed using the fundamental matrices (7) and
Kronecker products. It turns out that $C(0,2)$ is isomorphic to the
quaternion algebra $H$. Since all the others $C(a,b)$'s can be constructed
from the building blocks $C(2,0),$ $C(1,1)$ and $C(0,2)$, this means that
our connection between oriented matroid theory and $M(2,R)$ also establishes
an interesting link with the Clifford algebra structure (see Ref. [15].and
references therein).

Let us make some final remarks. The above links also apply to the subgroup $%
SL(2,R)$ which is the main object in 2t physics. In this case it is known
that noncommutative field theory of 2t physics [18-20] (see also Ref. [26])
contains a fundamental gauge symmetry principle based on the noncommutative
group $U_{\star }(1,1)$. This approach originates from the observation that
a world line theory admits a Lie algebra $sl_{\star }(2,R)$ gauge symmetry
acting on phase space [18]. In this context, consider the coordinates $q$
and $p$ in the phase-space. The Poisson bracket

\begin{equation}
\{f,g\}=\frac{\partial f}{\partial q^{a}}\frac{\partial g}{\partial p_{a}}-%
\frac{\partial f}{\partial p_{a}}\frac{\partial g}{\partial q^{a}},
\label{33}
\end{equation}%
can be written as

\begin{equation}
\{f,g\}=\varepsilon _{ij}\eta ^{ab}\frac{\partial f}{\partial q_{i}^{a}}%
\frac{\partial g}{\partial q_{j}^{b}}.  \label{34}
\end{equation}%
where $q_{1}^{a}\equiv q^{a}$ and $q_{2}^{a}\equiv p^{a}$, with $a$ and $b$
running from $1$ to $n$. It worth mentioning that the expression (34) is
very similar to the the definition of a chirotope (see Ref. [8] and
references therein).

Recently, a generalization of (34) was proposed [27], namely

\begin{equation}
\{f,g\}=(g_{ij}\Omega ^{ab}+\varepsilon _{ij}\eta ^{ab})\frac{\partial f}{%
\partial q_{i}^{a}}\frac{\partial g}{\partial q_{j}^{b}}.  \label{35}
\end{equation}%
Here, $\Omega ^{ab}$ is skew-simplectic form defined in even dimensions. In
particular, in four dimensions $\Omega _{ab}$ can be chosen as%
\begin{equation}
\Omega _{ab}=\left( 
\begin{array}{cccc}
0 & -1 & 0 & 0 \\ 
1 & 0 & 0 & 0 \\ 
0 & 0 & 0 & -1 \\ 
0 & 0 & 1 & 0%
\end{array}%
\right) .  \label{36}
\end{equation}%
Here, by choosing $\eta _{ab}=diag(-1,1,-1,1)$ we make contact with $(2+2)$%
-dimensions which is the minimal 2t physics theory (see Refs. [28--29]).

Let us write the factor in (35) $g_{ij}\Omega ^{ab}+\varepsilon _{ij}\eta
^{ab}$ in the form 
\begin{equation}
\mathbf{g}_{ij}^{\prime }+\varepsilon _{ij}\mathbf{\eta ,}  \label{37}
\end{equation}%
with $\mathbf{g}_{ij}^{\prime }=g_{ij}\mathbf{\Omega }$. We recognize in
(37) the typical form (18) for a complex structure. This proves that
oriented matroid theory is also connected not only with $(2+2)$-physics but
also with noncommutative geometry.

An alternative connection \ with 2t physics can be obtained by considering
the signature $\eta _{ab}=diag(1,1,-1,-1)$, and its associated metric:%
\begin{equation}
ds^{2}=(dx^{1})^{2}+(dx^{2})^{2}-(dx^{3})^{2}-(dx^{4})^{2}.  \label{38}
\end{equation}%
In fact, by defining \bigskip\ the matrix%
\begin{equation}
x^{ij}=%
\begin{pmatrix}
x^{1} & x^{3} \\ 
x^{4} & x^{2}%
\end{pmatrix}%
,  \label{39}
\end{equation}%
it can be seen that (38) can be expressed as%
\begin{equation}
ds^{2}=dx^{ij}dx^{kl}\eta _{ik}\eta _{jl},  \label{40}
\end{equation}%
where the indices $i,j,k,l$ run from 1 to 2 as before, and $\eta _{ij}$
stands for the third matrix defined in (7), namely $\eta _{ij}\equiv \left( 
\begin{array}{cc}
1 & 0 \\ 
0 & -1%
\end{array}%
\right) $. As before,noticing that in (38) the "spatial" coordinates $x^{1}$%
, $x^{2}$ are the elements of the main diagonal and the "time" coordinates $%
x^{3}$, $x^{4}$ corresponds to the main skew diagonal in (39), $x^{ij}$ can
be written in terms of the bases (7) as follows:%
\begin{equation}
x^{ij}=X\delta ^{ij}+S\varepsilon ^{ij}+Y\eta ^{ij}+T\lambda ^{ij};
\label{41}
\end{equation}%
where we used the definitions%
\begin{equation}
\begin{array}{ccc}
X=\frac{1}{2}(x^{1}+x^{2}), &  & S=\frac{1}{2}(x^{4}-x^{3}), \\ 
&  &  \\ 
Y=\frac{1}{2}(x^{1}-x^{2}), &  & T=-\frac{1}{2}(x^{3}+x^{4}),%
\end{array}
\label{42}
\end{equation}%
and considered the notation $\varepsilon ^{ij}=\varepsilon _{kl}\eta
^{ik}\eta ^{jl}$ and $\lambda ^{ij}=\lambda _{kl}\eta ^{ik}\eta ^{jl}$,
where $\eta ^{ij}$ is the inverse flat 1+1 metric, and has the same
components as $\eta _{ij}$.

\bigskip Finally, consider the three index object $\eta _{ijk}$ with
components 
\begin{equation}
\eta _{1ij}=\delta _{ij};\qquad \eta _{2ij}=\varepsilon _{ij}.  \label{43}
\end{equation}%
From these expressions and (7) it can be checked that $\eta _{ijk}$
automatically satisfies%
\begin{equation}
\eta _{ij1}=\eta _{ij};\qquad \eta _{ij2}=\lambda _{ij}.  \label{44}
\end{equation}%
Therefore $\eta _{ijk}$ has the remarkable property of containing all the
matrices in (7). This means that an arbitrary matrix $\Omega _{ij}$ can be
written as%
\begin{equation}
\Omega _{ij}=x^{k}\eta _{kij}+y^{k}\eta _{ijk},  \label{45}
\end{equation}%
where $x^{1}=x$, $x^{2}=y$ and $y^{1}=r$, $y^{2}=s$. Here, $x,y,r$ and $s$
are defined in (5). Observe that $\eta _{ijk}=\eta _{jik}$, but $\eta
_{kij}\neq \eta _{kji}$. It is worth mentioning that a similar structure was
proposed in Ref. [30] in the context of nonsymmetric gravity [31].

The inverse $\eta ^{ijk}$ of $\eta _{ijk}$ can be defined by the relation%
\begin{equation}
\eta ^{ijk}\eta _{ijl}=2\delta _{l}^{k},  \label{46}
\end{equation}%
or%
\begin{equation}
\eta ^{kij}\eta _{lij}=2\delta _{l}^{k}.  \label{47}
\end{equation}%
Explicity, we obtain the components 
\begin{equation}
\eta ^{1ij}=\delta ^{ij};\quad \eta ^{2ij}=-\varepsilon ^{ij};\quad \eta
^{ij1}=\eta ^{ij};\quad \eta ^{ij2}=-\lambda ^{ij}.  \label{48}
\end{equation}

Traditionally, starting with a flat space described by the metric $\eta
_{ij} $, one may introduce a curved metric $g_{\mu \nu }=$ $e_{\mu
}^{i}e_{\nu }^{j}\eta _{ij}$ via the \textit{zweibeins} $e_{\mu }^{i}$. So,
this motivate us to introduce the three-index curved metric 
\begin{equation}
g_{\mu \nu \lambda }=e_{\mu }^{i}e_{\nu }^{j}e_{\lambda }^{k}\eta _{ijk}.
\label{49}
\end{equation}%
It seems very interesting to try to develop a gravitational theory based in $%
g_{\mu \nu \lambda }$, for at least two reasons. First, the $\eta _{ijk}$
contains the four basic matrices (7), which we proved are linked to matroid
theory. Therefore this establishes a bridge between matroids and $g_{\mu \nu
\lambda }$. Thus, a gravitational theory based in $g_{\mu \nu \lambda }$ may
provide an alternative gravitoid theory (see Ref. [4]). Secondly, since the
matrices (7) are also linked to Clifford algebras, such a gravitational
theory may determine spin structures, which are necessary for supersymmetric
scenarios. These and another related developments will be reported elsewhere
[32].
\bigskip\ 
\bigskip\ 
\begin{center}
\textbf{Acknowledgments}
\end{center}

J. A. Nieto would like to thank to M. C. Mar\'{\i}n for helpful comments.
This work was partially supported by PROFAPI 2007 and PIFI 3.3.

\bigskip\

\end{document}